\documentstyle[12pt,epsf]{article}
\begin{document}
\begin{titlepage}
\hskip 11cm \vbox{\hbox{BUDKERINP/96-92}\hbox{December 1996}}
\vskip 0.3cm
\centerline{\bf GLUON PAIR PRODUCTION}
\centerline{\bf IN THE QUASI-MULTI-REGGE KINEMATICS$^{~\ast}$}
\vskip 0.8cm
\centerline{  V.S. Fadin$^{\dagger}$}
\vskip .1cm
\centerline{\sl Budker Institute for Nuclear Physics}
\centerline{\sl and Novosibirsk State University, 630090 Novosibirsk,
Russia}
\vskip .4cm
\centerline{  M.I. Kotsky$^{\dagger}$}
\vskip .1cm
\centerline{\sl Budker Institute for Nuclear Physics}
\centerline{\sl  630090 Novosibirsk, Russia}
\vskip .4cm
\centerline{  L.N. Lipatov$^{\ddagger}$}
\vskip .1cm
\centerline{\sl Petersburg Nuclear Physics Institute}
\centerline{\sl Gatchina, 188350 St.Petersburg, Russia}
\vskip 0.8cm
\begin{abstract}
To find the region of applicability of the leading log(1/x) approximation
for parton distributions in the small x
region and to fix the argument of the QCD running coupling it is
necessary to know radiative corrections to the kernel of the BFKL equation.
The next-to-leading corrections to the BFKL kernel are expressed
in terms of the two-loop correction to the gluon
Regge trajectory, one-loop correction
to the Reggeon-Reggeon-gluon  vertex,  and contributions from
two-gluon and quark-antiquark production in the quasi-multi-Regge
kinematics. We calculate differential and total cross sections of the
two gluon production. Differential cross section can be applied
for description of two jet production in the quasi-multi-Regge
kinematics; the total cross section defines corresponding correction to the
BFKL kernel. To escape the infrared divergencies we use dimensional
regularization of the Feynman integrals.
\end{abstract}
\vskip .5cm
\hrule
\vskip.3cm
\noindent

\noindent
$^{\ast}${\it Work supported by the Russian Fund of Basic Researches,
grants 95-02-04609 and 96-02-19299a.}
\vfill
\vskip .2cm
$ \begin{array}{ll}
^{\dagger}\mbox{{\it email address:}} &
 \mbox{FADIN, KOTSKY~@INP.NSK.SU}\\
\end{array}
$

$ \begin{array}{ll}
^{\ddagger}\mbox{{\it email address:}} &
  \mbox{LIPATOV@THD.PNPI.SPB.RU}
\end{array}
$
\vfill
\end{titlepage}

{\bf 1. Introduction}

The problem of calculation of the parton distributions  in
the small $x$ region can be turned into calculation of the kernel of the
Bethe-Salpeter type equation for the $t$-channel partial amplitude with
the vacuum quantum numbers \cite{1}. This equation is known now as BFKL-
equation. In the leading logarithmic approximation (LLA), which means summation
of all terms of the type $[\alpha_s\ln(1/x)]^n$, this kernel was found many
years ago \cite{1}. Now the results of LLA are widely known and used for
description of experimental data.

The LLA gives a power growth of cross sections with c.m.s. energy
$\sqrt{s}$. In terms of parton distributions this means a fast increase of
the gluon density $g(x,Q^2)$ in the small $x$ region:
$$
g(x,Q^2) \sim x^{-j_0}\ ,
$$
where $j_0 = 1 + \omega_0$ is the LLA position of the singularity of the
partial amplitude with the vacuum quantum numbers in the $t$-channel \cite{1}:
$$
\omega_0 = \frac{4\alpha_s}{\pi}N\ln 2,
$$
with $N = 3$ for QCD. Such behaviour contradcts the unitarity and therefore
the LLA can not be applied at asymptotically small x. Unitarity constraints
for scattering amplitudes with vacuum quantum numbers in $t$-channel don't
work in this approximation and, as a result, the Froissart bound
$\sigma_{tot} < const(\ln s)^2$ is violated. Nevertheless, in the region
of parameters accessible for modern experiments the observed behaviour
of the structure functions is consistent with LLA results \cite{2}, and
we will not discuss here the unitarization problem, which appear at
asymptotically large energies.

From practical point of view it seems more important to determine the region
of applicability of LLA. Besides that, the dependence of the QCD running
coupling  $\alpha_s$ on virtuality is beyond of the accuracy of the LLA.
It diminishes the predictive power of the LLA, because numerical results of
this approximation can be strongly modified by changing a scale of virtuality.
All these uncertainties of LLA predictions can be removed and the region
of applicability of these predictions can be found using radiative
corrections.

As it is clear from above discussion the problem of calculation of the
next-to-leading corrections to the BFKL kernel is very important now.
These corrections are expressed \cite{3}
in terms of the two-loop correction to the gluon
Regge trajectory, one-loop correction
to the Reggeon-Reggeon-gluon (RRG) vertex,  and contrbutions from two-gluon
and quark-antiquark production in the quasi-multi-Regge
kinematics (QMRK), which, in turn, are expressed in terms of the Reggeon-
Reggeon-two-gluon (RRGG) and Reggeon-Reggeon-quark-antiquark (RRq$\bar q$)
vertices.
Corrections to the RRG vertex and to the Reggeized gluon
trajectory were calculated (see Refs. \cite{4} and \cite{5}
correspondingly), after that calculation
of the contributions from two-gluon and
quark-antiquark production in the QMRK became the most urgent
problem.

Investigation of these contributions was started
in \cite{3}, where the two-gluon production amplitude in the QMRK, and,
correspondingly, the RRGG vertex, was
found. The next important step was done in \cite{6}, where
the two-gluon and quark-antiquark production amplitudes in the QMRK
were simplified  using the helicity representation, the corresponding
next to leading
contributions to the BFKL kernel were expressed in
terms of the integrals from the squares of these helicity amplitudes over
transverse and longitudinal momenta of produced particles,
all infrared divergencies were extracted from these expressions in an
explicit form and it
was demonstrated for the fermion contribution that these
divergencies cancel with the analogous divergencies from the virtual
corrections to the BFKL equation. Recently resummation formulas for the
quark-antiquark production contribution to the BFKL kenel was derived
in \cite{8}.

In this paper we calculate differential and total cross sections of the
two gluon production in QMRK. Differential cross section can be applied
for description of two jet production in the QMRK;
the total cross section defines corresponding contribution in the
next-to-leading BFKL kernel. To escape the infrared divergencies we
use dimensional regularization of the Feynman integrals. All divergencies
cancel in the total expression for the
next-to-leading BFKL kernel. This cancellation was demonstrated in explicit
form in Refs. \cite{6}, \cite{7}.

The paper is organized as follows. In the next
Section we discuss kinematics and method of calculations. In Section 3 and
4 we calculate the differential and total cross sections respectively. The
final result for the total cross section, needed to get the corresponding
correction to the BFKL kernel, is presented and briefly discussed in the
Section 4.

{\bf 2. The amplitudes in QMRK}

  Let us consider the two-gluon production in the QMRK in
gluon-gluon scattering at high energies in the Born approximation
(see Fig. 1). The corresponding contribution to the imaginary part of the
elastic gluon-gluon scattering amplitude $A_{AB}^{AB}$ at zero momentum
transfer is proportional to the cross section $\sigma_1$ of the process
Fig. 1.

\begin{figure}
\begin{center}
{ \parbox[t]{5cm}{\epsfysize 5cm \epsffile{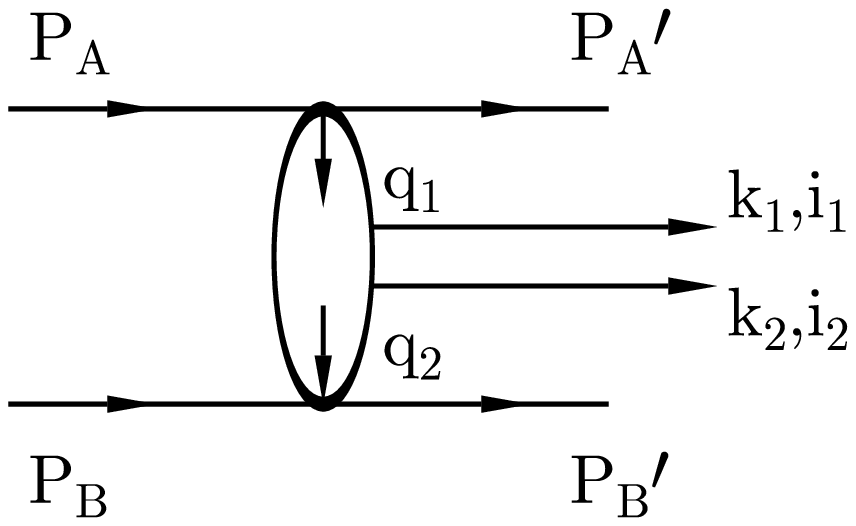}}}
\end{center}
\caption{}
\label{fig1}
\end{figure}

  We will use the Sudakov decomposition for the momenta of the final gluons:
\begin{equation}\label{1}
k_i = \beta_ip_A + \alpha_ip_B + {k_i}_{\perp}\ ,\ \ \ s\alpha_i\beta_i =
-{k_i}_{\perp}^2,
\end{equation}
where $s = (p_A + p_B)^2$ is the total  c.m.s. energy squared, which is
supposed to be tending to infinity, and we took into account the
on-mass-shellness of these particles. Then the QMRK means that all ${k_i}_{\perp}$
are fixed and
\begin{equation}\label{2}
\beta_{A^{\prime}} \gg \beta_1 \sim \beta_2 \gg \beta_{B^{\prime}}.
\end{equation}
The squared invariant mass of the gluons 1 and 2, $\kappa = (k_1 + k_2)^2$,
in the QMRK has the same order of magnitude as $|{k_i}_{\perp}^2|$. The amplitude
for this process was calculated in Ref.\cite{3} and can be presented in
the following factorized form:
\begin{equation}\label{3}
A = 2sg^2 \delta_{\lambda_{A^{\prime}},\lambda_A} T^c_{A^{\prime}A} \frac
{1}{{q_1}_{\perp}^2} \gamma_{cd}^{i_1i_2}(q_1,q_2) \delta_{\lambda_{B^
{\prime}},\lambda_B} T^d_{B^{\prime}B} \frac{1}{{q_2}_{\perp}^2},
\end{equation}
where $T^k_{ij}$ are the generators of the $SU(N)$ colour group in the
adjoint representation, $\lambda_i$ are the helicities of the corresponding
particles and $g$ is the gauge coupling constant ($\alpha_s = g^2/4\pi$).
The expression for the effective RRGG vertex
$\gamma_{cd}^{i_1i_2}(q_1,q_2)$ can be
found in Ref.\cite{3} and we don't write it here. Let us note that, although
QMRK means $\beta_1 \sim \beta_2$, the amplitude (\ref{3}) is correct
also in the multi-Regge kinematics (MRK), when $\beta_1 \gg \beta_2$ (or
$\beta_2 \gg \beta_1$).

The cross section $d\sigma_1$ for the process, pictured on Fig. 1, can be
presented in the form (we average over colours and spins of the initial
particles and sum over the same quantum numbers of the final particles):
$$
d\sigma_1 = \frac{N^2\alpha_s^2}{(N^2-1)(2\pi)^{2D-5}}\frac{d^{D-2}{q_1}_{\perp
}d^{D-2}{\Delta}_{\perp}}{({q_1}_{\perp}^2(q_1 - \Delta)_{\perp}^2)^2}\frac
{d\beta}{\beta}d\sigma,
$$
$$
d\sigma = \frac{1}{2!}(2\pi)^{D-1}\delta^{(D-2)}((k_1+k_2-\Delta)_{\perp})
\frac{dx}{x(1-x)}\frac{d^{D-2}{k_1}_{\perp}}{(2\pi)^{D-1}}\frac{d^{D-2}{k_2}_
{\perp}}{(2\pi)^{D-1}}
$$
\begin{equation}\label{4}
\times P_0^{cc^{\prime}dd^{\prime}}\sum_{\lambda_1,\lambda_2}\gamma_{cd}^{i_1i_
2}(q_1,q_2){\gamma^*}_{c^{\prime}d^{\prime}}^{i_1i_2}(q_1,q_2),
\end{equation}
where
\begin{equation}\label{5}
\beta = \beta_1 + \beta_2,\ \ \ x = \frac{\beta_1}{\beta} = 1 - \frac{\beta_2}
{\beta},
\end{equation}
and
\begin{equation}\label{6}
P_0^{cc^{\prime}dd^{\prime}} = \frac{\delta_{cc^{\prime}}\delta_{dd^{\prime}}}
{(N^2-1)}
\end{equation}
is the projector on the colour singlet state in the $t$- channel of the
elastic scattering amplitude. D here is the space-time dimension, different
from 4 to regularize the cross section, having infrared and collinear
divergencies.

$d\sigma$ can be named as two-gluon production cross section in QMRK and
corresponding correction to the BFKL kernel is equal to (see \cite{6}):
\begin{equation}\label{7}
K^{(2)}_{gluons} = \frac{\mu^{4-D}}{4(2\pi)^{D-1}}\frac{\sigma_{tot}(q_1,q_2)}
{{q_1}_{\perp}^2{q_2}_{\perp}^2}.
\end{equation}
As it was shown in Ref. \cite{6} the cross section $d\sigma$ can be expressed
through the variable $x$ and transverse momenta of the produced gluons in the
following way:
$$
d\sigma = 8g^4\frac{dx}{x(1-x)}\frac{d^{D-2}{k_1}_{\perp}}{(2\pi)^{D-1}}
P_0^{cc^{\prime}dd^{\prime}}\biggl[ \biggl\{ H_1^{cc^{\prime}dd^{\prime}}
a^{\mu\nu}a_{\mu\nu} + H_2^{cc^{\prime}dd^{\prime}}\tilde{a}^{\mu\nu}
\tilde{a}_{\nu\mu}( x \leftrightarrow 1-x, {k_1}_{\perp} \leftrightarrow
{k_2}_{\perp} ) \biggr\}
$$
\begin{equation}\label{8}
+ \biggl\{ x \leftrightarrow 1-x, {k_1}_{\perp}
\leftrightarrow {k_2}_{\perp} \biggr\} \biggr],
\end{equation}
where
$$
H_1^{cc^{\prime}dd^{\prime}} = T_{ci}^{i_1}T_{id}^{i_2}T_{c^{\prime}j}^{i_1}T_
{jd^{\prime}}^{i_2},\ \ \ P_0^{cc^{\prime}dd^{\prime}}H_1^{cc^{\prime}dd^
{\prime}} = N^2,
$$
\begin{equation}\label{9}
H_2^{cc^{\prime}dd^{\prime}} = T_{ci}^{i_1}T_{id}^{i_2}T_{c^{\prime}j}^{i_2}T_
{jd^{\prime}}^{i_1},\ \ \ P_0^{cc^{\prime}dd^{\prime}}H_2^{cc^{\prime}dd^
{\prime}} = N^2/2
\end{equation}
are the different colour factors,
\begin{equation}\label{10}
\tilde{a}^{\mu\nu} = \left( g_{\perp\perp}^{\mu\rho} - \frac{{k_1}_{\perp}^
{\mu}{k_1}_{\perp}^{\rho}}{{k_1}_{\perp}^2} \right){a_{\rho}}^{\nu},
\end{equation}
and
$$
a^{\mu\nu} = \frac{(q_1-k_1)_{\perp}^{\mu}(q_1-k_1)_{\perp}^{\nu}}{t} -
\frac{(q_1-k_1)_{\perp}^{\mu}}{\kappa}\left( k_1 - \frac{x}{1-x}k_2 \right)_
{\perp}^{\nu} + \left( k_2 - \frac{x{k_2}_{\perp}^2}{(1-x){k_1}_{\perp}^2}k_1
\right)_{\perp}^{\mu}
$$
$$
\times \frac{(q_1-k_1)_{\perp}^{\nu}}{\kappa} - \frac{{k_1}_{\perp}^{\mu}{k_1}
_{\perp}^{\nu}}{\kappa}\frac{xt_2}{{k_1}_{\perp}^2} - \frac{{k_2}_{\perp}^{\mu}
{k_2}_{\perp}^{\nu}}{\kappa}\frac{xt_1}{z} - \frac{{k_1}_{\perp}^{\mu}{k_2}_
{\perp}^{\nu}}{\kappa}\left( 1 - \frac{xt}{(1-x){k_1}_{\perp}^2} \right) +
\frac{{k_2}_{\perp}^{\mu}{k_1}_{\perp}^{\nu}}{\kappa}
$$
\begin{equation}\label{11}
- \frac{1}{2}g_{\perp\perp}^{\mu\nu}\left( 1 + \frac{t}{\kappa} - \frac{(1-x)
{k_1}_{\perp}^2}{xt} + \frac{x{k_2}_{\perp}^2}{(1-x)\kappa} - \frac{(1-x){k_1}
_{\perp}^2}{x\kappa} - \frac{xt_1{k_2}_{\perp}^2}{\kappa z} - \frac{xt_2}
{\kappa} \right).
\end{equation}
In the above expressions we used the notations ($\kappa$ is the invariant
mass of the produced pair as earlier)
$$
\kappa = -\frac{((1-x)k_1 - xk_2)_{\perp}^2}{x(1-x)},\ \ \ z = (1-x){k_1}_
{\perp}^2 + x{k_2}_{\perp}^2,
$$
$$
t = \frac{1}{x}\left( (k_1 - xq_1)_{\perp}^2 + x(1-x){q_1}_{\perp}^2 \right),
\ \ \ {k_2}_{\perp} = (\Delta - k_1)_{\perp},
$$
\begin{equation}\label{12}
t_1 = {q_1}_{\perp}^2,\ \ \ t_2 = {q_2}_{\perp}^2,
\end{equation}
and $g_{\perp\perp}^{\mu\nu}$ is the metrical tensor in the transverse
subspace
\begin{equation}\label{13}
g_{\perp\perp}^{\mu\nu} = g^{\mu\nu} - \frac{p_A^{\mu}p_B^{\nu} + p_B^{\mu}p_A^
{\nu}}{(p_Ap_B)}.
\end{equation}

In the next sections we calculate the differential cross section $d\sigma
/dx$ and the total cross section. Since all vectors, entering to
(\ref{11}), are transverse we will omit below the sign of the transversality.

{\bf 3. The differential cross section.}

  After rather long calculation we get for the convolutions of the tensors,
entering to (\ref{8})
$$
a^{\mu\nu}a_{\mu\nu} = \frac{1}{\kappa t} \biggl[ 2xq_1^2q_2^2 +
\left( \frac{D-2}{4} \right)x(1-x)\biggl( 2(1-x)q_1^2(\Delta^2-q_2^2) -
x\left( (q_1^2)^2 + (\Delta^2-q_2^2)^2 \right) \biggr) -
$$
$$
- (D-2)(1-x)^2q_1^2(q_1k_1) \biggr] + \frac{x(1-x)^2q_1^2}{2\kappa z}\biggl[
(D-2)(1-x)q_1^2 - (D-4)x(\Delta^2-q_2^2) \biggr] +
$$
$$
+ \biggl( \frac{x(1-x)q_1^2}{z} \biggr)^2\biggl[ \left( \frac{D-2}{4} \right) -
(D-1)x(1-x) \biggr] + \left( \frac{D-2}{4} \right)(1-4x)\biggl( \frac{(1-x)q_1^
2}{t} \biggr)^2 -
$$
$$
- \frac{xq_1^2q_2^2}{2(1-x)\kappa k_1^2} + \frac{x^2q_1^2q_2^2}{2k_1^2z} -
\frac{x(q_1^2)^2q_2^2}{2\kappa tk_1^2} - \frac{2xq_1^2}{\kappa zt}\biggl[
\left( \frac{q_2^2}{2} \right)^2 -
(1-x)q_1^2q_2^2 + \left( \frac{D-2}{4} \right)(1-x)^2\left( q_1^2 \right)^2 +
$$
$$
+ (1-x)\left( 2q_2^2 - (D-2)(1-x)q_1^2 \right)(k_1q_1) + (D-2)(1-x)^2(k_1q_1)^
2 \biggr] +
$$
\begin{equation}\label{14}
+ (D-2)\biggl( \frac{q_1^2-x\Delta^2-2k_1q_2}{\kappa t} \biggr)\biggl(
\frac{(1-x)(k_1q_1)^2}{t} + \frac{\left( q_1(k_1-x\Delta) \right)^2}{\kappa}
\biggr) + ...\ \ ,
\end{equation}
and
$$
2\tilde{a}^{\mu\nu}\tilde{a}_{\nu\mu}( x \leftrightarrow 1-x, {k_1}
\leftrightarrow {k_2} ) = \biggl\{ -a^{\mu\nu}a_{\mu\nu} + \frac{q_1^2q_2^2}
{2k_1^2k_2^2} + \frac{1}{2t\tilde{t}}\biggl[ -2q_1^2q_2^2 + \left(\frac{D-2}{4}
\right)x(1-x)
$$
$$
\times \left( (\Delta^2 - q_2^2)^2 + (q_1^2)^2 \right) \biggr] + \left( \frac
{D-2}{4} \right)(1-4x)\biggl( \frac{(1-x)q_1^2}{t} \biggr)^2 + (D-2)\frac
{(1-x)(k_1q_1)^2}{\kappa t^2}
$$
$$
\times (q_1^2 - x\Delta^2 - 2k_1q_2) + \frac{(1-x)(2k_1q_1 - q_1^2)q_1^2q_2^2}
{2tk_1^2k_2^2} - (D-2)\frac{(1-x)(k_1q_1)^2(q_2^2 + x\Delta^2 + 2k_1q_2)}
{\kappa t\tilde{t}}
$$
\begin{equation}\label{15}
+ \frac{(q_1^2)^2(q_2^2)^2}{4t\tilde{t}k_1^2k_2^2} \biggr\} + \biggl\{ x
\leftrightarrow 1-x, \ \ k_1 \leftrightarrow k_2 \biggr\} + ...\ \ ,
\end{equation}
where ... means the terms, vanishing at integration over $k_1$ and we
introduced the notation
$$
\tilde{t} = t(x \leftrightarrow 1-x, \ \ k_1 \leftrightarrow k_2) = \frac{1}
{1-x}\left( (k_2 - (1-x)q_1)^2 + x(1-x)q_1^2 \right) = \frac{1}{1-x}
$$
\begin{equation}\label{16}
\times \left( (k_1 + (q_2 - xq_1))^2 + x(1-x)q_1^2 \right) = q_1^2 + q_2^2 -
\kappa - t = q_1^2 + q_2^2 - \Delta^2 + \frac{z}{x(1-x)} - t.
\end{equation}

Using the relations (\ref{8})-(\ref{12}) and (\ref{14})-(\ref{16}) one can
get
\begin{equation}\label{17}
\mu^{4-D}\frac{x(1-x)}{4g^4N^2}\frac{d\sigma}{dx} = I + I( x \leftrightarrow
1-x ),
\end{equation}
$$
I = \int \mu^{4-D} \frac{d^{D-2}k_1}{(2\pi)^{D-1}} \biggl\{ \frac{1}{\kappa t}
\biggl[ 2xq_1^2q_2^2 + \left( \frac{D-2}{4} \right)x(1-x)\biggl( 2(1-x)q_1^2
(\Delta^2 - q_2^2) - x(q_1^2)^2
$$
$$
- x(\Delta^2-q_2^2)^2 \biggr) - (D-2)(1-x)^2q_1^2(k_1q_1) \biggr] + 2(D-2)\frac
{(1-x)(k_1q_1)^2(q_1^2 - x\Delta^2 - 2k_1q_2)}{\kappa t^2}
$$
$$
+ (D-2)\frac{((k_1 - x\Delta)q_1)^2(q_1^2 - x\Delta^2 - 2k_1q_2)}{\kappa^2t} -
(D-2)\frac{(1-x)(k_1q_1)^2(q_2^2 + x\Delta^2 + 2k_1q_2)}{\kappa t\tilde{t}}
$$
$$
- \frac{2xq_1^2}{\kappa zt}\biggl[ \left( \frac{q_2^2}{2} \right)^2 - (1-x)q_1^
2q_2^2 + \left( \frac{D-2}{4} \right)(1-x)^2(q_1^2)^2 + (1-x)\left( 2q_2^2 -
(D-2)(1-x)q_1^2 \right)
$$
$$
\times (k_1q_1) + (D-2)(1-x)^2(k_1q_1)^2 \biggr] + \frac{1}{2t\tilde{t}}\biggl[
-2q_1^2q_2^2 + \left( \frac{D-2}{4} \right)x(1-x)\left( (\Delta^2 - q_2^2)^2 +
(q_1^2)^2 \right) \biggr]
$$
$$
+ \frac{x(1-x)^2q_1^2}{2\kappa z}\biggl[ (D-2)(1-x)q_1^2 - (D-4)x(\Delta^2-q_2^
2) \biggr] + \biggl( \frac{x(1-x)q_1^2}{z} \biggr)^2\biggl[ \left( \frac{D-2}
{4} \right) - (D-1)
$$
$$
\times x(1-x) \biggr] + \left( \frac{D-2}{2} \right)(1-4x)\biggl( \frac{(1-x)q_
1^2}{t} \biggr)^2 + \frac{(1-x)(2k_1q_1 - q_1^2)q_1^2q_2^2}{2tk_1^2k_2^2} -
\frac{xq_2^2(q_1^2)^2}{2\kappa tk_1^2}
$$
\begin{equation}\label{18}
+ \frac{(q_1^2)^2(q_2^2)^2}{4t\tilde{t}k_1^2k_2^2} - \frac{xq_1^2q_2^2}{2(1-x)
\kappa k_1^2} + \frac{x^2q_1^2q_2^2}{2k_1^2z} + \frac{q_1^2q_2^2}{2k_1^2k_2^2}
\biggr\}.
\end{equation}

Let us denote the contribution to $I$ of the term number $j$ in the integrand
of the above expression as $I_j$. All $I_j$ demonstrate convergence at large
$k_1$ and have only logarithmic infrared divergencies. Some of these integrals
can be calculated exactly, anothers - only in the form of the expansion in
$\epsilon = (D-4)/2$. Because the BFKL equation contains integration over
$\Delta$, we should calculate the cross
section $d\sigma/dx$ with an accuracy up to terms, giving, after integration
over $x$ and $\Delta$, nonvanishing in the physical limit $\epsilon
\rightarrow 0$ contributions. The results of calculation of the integrals
$I_j$ can be found in the Appendix I. Using this table for $I_j$ and Eqs.
(\ref{17}), (\ref{18}) one can get the differential cross section $d\sigma/dx$.

{\bf 4. The total cross section.}

To calculate the total cross section $\sigma_{tot}$ one can integrate
$d\sigma/dx$ over $x$, using Eqs. (\ref{17}), (\ref{18}) and the expressions
from Appendix I for the integrals $I_j$. But it is very hard way. The matter
is that after integration over $k_1$ the additional integrations over Feynman
variables appear. In the final expressions of Appendix I these additional
integrations have been performed at fixed $x$. But at calculation the total
cross section we can change the orders of integrations over $x$ and over
Feynman variables in the most convenient way so that integrations become
simpler. The such changes of the orders of the integrations strongly
simplify calculation of the total cross section.

Slightly rearranging the Eqs. (\ref{17}), (\ref{18}) we get
$$
\frac{\mu^{-2\epsilon}\sigma_{tot}}{8g^4N^2} = \int_{\delta_R}^{1-\delta_R}
\frac{dx}{x(1-x)} \int \mu^{-2\epsilon}\frac{d^{D-2}k_1}{(2\pi)^{D-1}} \biggl\{
(1 + \epsilon) (1-x)^2 \frac{4(k_1q_1)^2 + (1-4x)(q_1^2)^2}{t^2}
$$
$$
+ \frac{x(1-x)q_1^2}{\kappa z}\biggl[ 2q_2^2 +(1+\epsilon)\left( 2(1-x)(k_1q_1)
- x(1-x)q_1^2 - k_2^2 \right) - \epsilon x(1-x) (\Delta^2 - q_2^2) \biggr]
$$
$$
+ \left( \frac{x(1-x)q_1^2}{z} \right)^2\biggl[ \frac{(1 + \epsilon)}{2} - (3
+ 2\epsilon)x(1-x) \biggr] + \biggl( \frac{-xq_1^2q_2^2}{2(1-x)\kappa k_1^2}
+ \frac{x^2q_1^2q_2^2}{2k_1^2z} + \frac{xq_1^2q_2^2}{k_1^2k_2^2} \biggr)
$$
$$
+ \frac{1}{\kappa t}\biggl[ -2(1-x)q_1^2q_2^2 + (1 + \epsilon) (1-x) (q_1^2)^2
+ \frac{(1 + \epsilon)}{2}x(1-x)\left( 2(1-x)q_1^2(\Delta^2 - q_2^2) -
x(q_1^2)^2 \right.
$$
$$
\left. -x(\Delta^2 - q_2^2)^2 \right) -2(1 + \epsilon)(2-x)(1-x)q_1^2(k_1q_1)
+ 2(1 + \epsilon)(1-x)\left( (k_1q_1)^2 + ((k_1 - x\Delta)q_1)^2 \right)
$$
$$
+ (1 + \epsilon)q_1^2k_2^2 \biggr] - \frac{xq_1^2k_2^2\left( (1 + \epsilon)
k_2^2 -2q_2^2 \right)}{\kappa zt} - \frac{xq_1^2(q_2^2)^2}{2\kappa zt} - \frac
{xq_2^2(q_1^2)^2}{2\kappa tk_1^2} + \frac{(1-x)(2(k_1q_1) - q_1^2)q_1^2q_2^2}
{2tk_1^2k_2^2}
$$
\begin{equation}\label{19}
+ \frac{1}{2t\tilde{t}}\biggl[ -2q_1^2q_2^2 + \frac{(1 + \epsilon)}{2}x(1-x)
\left( (\Delta^2 - q_2^2)^2 + (q_1^2)^2 \right) - 4(1 + \epsilon)x(1-x)
(k_1q_1)^2 \biggr] + \frac{(q_1^2)^2(q_2^2)^2}{4t\tilde{t} k_1^2k_2^2}
\biggr\}.
\end{equation}
We integrate over $x$ here from $\delta_R$ to $(1-\delta_R)$ to remove
the contribution of the multi-Regge kinematics, which is accounted by
leading term in the BFKL kernel. In the total expression with account of
the next-to-leading terms for this kernel the $\delta_R$- dependence
disappear. Technically to get corresponding correction to the kernel we
should omit the term with $\ln(1/\delta_R)$ in the total cross section
at substitution $\sigma_{tot}$ in the expression for the BFKL kernel
(\ref{7}) and leave the leading term in the kernel the same as earlier.

Let us again denote contribution of the term number $i$ in the integrand
in the RHS Eq. (\ref{19}) as $J_i$ and calculate these integrals separately.
Let us note, that contrary to the case of the differential cross section,
the separate integrals $J_i$ contain "ultraviolet" divergencies at large
$k_1$, but these divergencies are regularized by the same $\epsilon$ as
infrared ones and, as it is clear from Eqs. (\ref{17}), (\ref{18}), they
must cancel in the final expression for $\sigma_{tot}$ after summation of
all $J_i$. Using the results of calculation of these integrals from Appendix
II we get for the total cross section:
$$
\frac{\mu^{-2\epsilon}\sigma_{tot}}{8g^4N^2} = \frac{\Gamma(1-\epsilon)}
{(4\pi)^{2+\epsilon}}\Biggl\{ \frac{{\vec q_1}^2{\vec q_2}^2}{{\vec \Delta}^2}
\left( \frac{{\vec \Delta}^2}{\mu^2} \right)^{\epsilon} \frac{2\Gamma^2(1 +
\epsilon)}{\epsilon\Gamma(1+2\epsilon)}\Biggl[ 4\ln(1/\delta_R) + \frac{1}
{\epsilon} + \psi(1) + \psi(1+\epsilon) - 2\psi(1+2\epsilon)
$$
$$
- \frac{(11+7\epsilon)}{2(1+2\epsilon)(3+2\epsilon)} \Biggr] - \frac{({\vec
q_1}^2+{\vec q_2}^2)(2{\vec q_1}^2{\vec q_2}^2-3({\vec q_1}{\vec q_2})^2)}
{8{\vec q_1}^2{\vec q_2}^2} - \Biggl( \frac{11}{3}\frac{{\vec q_1}^2{\vec
q_2}^2}{({\vec q_1}^2-{\vec q_2}^2)} + \frac{(2{\vec q_1}^2{\vec q_2}^2 -
3({\vec q_1}{\vec q_2})^2)}{16{\vec q_1}^2{\vec q_2}^2}
$$
$$
\times ({\vec q_1}^2-{\vec q_2}^2) \Biggr)\ln\left( \frac{{\vec q_1}^2}
{{\vec q_2}^2} \right) - \frac{2}{3}\frac{{\vec q_1}^2{\vec q_2}^2}{({\vec q_1}
^2-{\vec q_2}^2)^3}\Biggl[ \left( 1-\frac{2({\vec q_1}{\vec q_2})^2}{{\vec q_1}
^2{\vec q_2}^2} \right)\Biggl( {\vec q_1}^4-{\vec q_2}^4 - 2{\vec q_1}^2{\vec
q_2}^2\ln\left( \frac{{\vec q_1}^2}{{\vec q_2}^2} \right) \Biggr) + ({\vec q_1}
{\vec q_2})
$$
$$
\times \left( 2({\vec q_1}^2-{\vec q_2}^2)-({\vec q_1}^2+{\vec q_2}^2) \ln
\left( \frac{{\vec q_1}^2}{{\vec q_2}^2} \right) \right) \Biggr] - \Biggl[
4{\vec q_1}^2{\vec q_2}^2 + \frac{({\vec q_1}^2-{\vec q_2}^2)^2}{4} + (2{\vec
q_1}^2{\vec q_2}^2-3{\vec q_1}^4-3{\vec q_2}^4)
$$
$$
\times \frac{(2{\vec q_1}^2{\vec q_2}^2-({\vec q_1}{\vec q_2})^2)}{16{\vec
q_1}^2{\vec q_2}^2} \Biggr]\int_0^{\infty}\frac{dx}{({\vec q_1}^2+x^2{\vec
q_2}^2)}\ln\left| \frac{1+x}{1-x} \right| + \frac{2{\vec q_1}^2{\vec q_2}^2
({\vec \Delta}(\vec q_1 + \vec q_2))}{{\vec \Delta}^2(\vec q_1 + \vec q_2)^2}
\Biggl[ \ln\left( \frac{{\vec q_1}^2}{{\vec q_2}^2} \right)
$$
$$
\times \ln \left( \frac{{\vec q_1}^2{\vec q_2}^2}{({\vec q_1}^2+{\vec
q_2}^2)^2} \right) + L\left( 1-\frac{{\vec \Delta}^2}{{\vec q_2}^2} \right) -
L\left( 1-\frac{{\vec \Delta}^2}{{\vec q_1}^2} \right) + L\left( -\frac
{{\vec q_1}^2}{{\vec q_2}^2} \right) - L\left( -\frac{{\vec q_2}^2}{{\vec
q_1}^2} \right) \Biggr] + 2{\vec q_1}^2{\vec q_2}^2
$$
$$
\times \Biggl[ \Biggl( \int_0^1\frac{dt}{({\vec q_2}^2t^2-2({\vec q_1}{\vec
q_2})t + {\vec q_1}^2)}\Biggl( \frac{({\vec q_2}{\vec \Delta})}{{\vec
\Delta}^2} - \frac{{\vec q_2}^2({\vec \Delta}(\vec q_1 + \vec q_2))}{{\vec
\Delta}^2(\vec q_1 + \vec q_2)^2}(1+t) \Biggr) \ln \left( \frac{{\vec
q_2}^2t(1-t)}{{\vec q_1}^2(1-t)+{\vec \Delta}^2t} \right) \Biggl)
$$
\begin{equation}\label{20}
+ \Biggl( \vec q_1 \leftrightarrow -\vec q_2 \Biggr) \Biggr] \Biggr\},\
\ \ \ \ L(z) = \int_0^z \frac{dt}{t} \ln(1-t),
\end{equation}
where we passed to the $(D-2)$- dimensional Euclidian vectors in the
final expression, because the transverse momenta are spacelike, and
$\psi(z)$ here is the logarithmic derivative of the gamma function:
\begin{equation}\label{21}
\psi(z) = \frac{\Gamma^{\prime}(z)}{\Gamma(z)}.
\end{equation}
Eq. (\ref{20}) is correct with an accuracy up to terms, giving in the phisical
case $\epsilon \rightarrow 0$ non zero contributions at subsequent integration
over $\vec \Delta$. Throwing away the term with $\ln(1/\delta_R)$ in the RHS
of Eq. (\ref{20}) and substituting then (\ref{20}) to (\ref{7}) we get the
correction to the BFKL kernel, connected with the two-gluon production in the
QMRK. The terms in this correction, which are singular at $\epsilon
\rightarrow 0$, cancel in the BFKL equation with corresponding terms in the
corrections to RRG vertex and to the gluon Regge trajectory, as it was shown
in Ref. \cite{7}.

\newpage
{\bf Appendix I.}

  Here we calculate the transverse momentum integrals appearing in Eq. (\ref
{18}). Because the transverse momenta are spacelike we pass to $(D-2)$-
dimensional Euclidian vectors in the final expressions. Let us remind, that
the expressions for the invariants $\kappa, z, t, \tilde{t}$ in terms of the
integration variables are given by Eqs. (\ref{12}), (\ref{16}) and
$$
\epsilon = (D-4)/2.
$$
Then we get
$$
I_1 = \int \mu^{-2\epsilon}\frac{d^{D-2}k_1}{(2\pi)^{D-1}} \frac{1}{\kappa t}
\biggl[ 2xq_1^2q_2^2 + \left( \frac{D-2}{4} \right)x(1-x)\biggl( 2(1-x)q_1^2
(\Delta^2 - q_2^2) - x(q_1^2)^2
$$
$$
- x(\Delta^2-q_2^2)^2 \biggr) - (D-2)(1-x)^2q_1^2(k_1q_1) \biggr] = \frac
{2\Gamma(1-\epsilon)}{(4\pi)^{2 + \epsilon}}(-x^2(1-x))\left( \frac{x(1-x)
{\vec q_1}^2}{\mu^2} \right)^{\epsilon}
$$
$$
\times \biggl\{ \frac{1}{((1-x){\vec q_1}^2 + x{\vec q_2}^2)} \left[ 2{\vec
q_1}^2{\vec q_2}^2 + \frac{(1 + \epsilon)}{2}(1-x)\left( (x-2){\vec q_1}^4 -
x({\vec \Delta}^2 - {\vec q_2}^2)^2 \right) \right] \biggl( \frac{1}{\epsilon}
$$
\begin{equation}\label{AI1}
+ 2\ln\left( \frac{(1-x){\vec q_1}^2 + x{\vec q_2}^2}{(1-x){\vec q_1}^2}
\right) \biggr) - \frac{(1-x)^2}{x}\frac{(2{\vec q_1}{\vec q_2}){\vec q_1}^2}
{{\vec q_2}^2}\ln\left( \frac{(1-x){\vec q_1}^2 + x{\vec q_2}^2}{(1-x){\vec
q_1}^2} \right) \biggr\}.
\end{equation}

$$
I_2 = \int \mu^{-2\epsilon}\frac{d^{D-2}k_1}{(2\pi)^{D-1}} 2(D-2)\frac{(1-x)
(k_1q_1)^2(q_1^2 - x\Delta^2 - 2k_1q_2)}{\kappa t^2} = \frac{2\Gamma(1 -
\epsilon)}{(4\pi)^{2 + \epsilon}}(-4x^2(1-x)^2)
$$
$$
\times \biggl\{ \frac{x({\vec q_1}{\vec \Delta})^2}{((1-x){\vec q_1}^2 +
x{\vec q_2}^2)} \biggl( \frac{1}{\epsilon} + 1 + \ln\left( \frac{x((1-x){\vec
q_1}^2 + x{\vec q_2}^2)^2}{(1-x){\vec q_1}^2\mu^2} \right) \biggr) + \left(
({\vec q_1}{\vec q_2})^2 - \frac{{\vec q_1}^2{\vec q_2}^2}{2} \right)
$$
$$
\times \left( \frac{(1-x){\vec q_1}^2 + x{\vec q_2}^2}{x{\vec q_2}^4} \right)
\ln\left( \frac{(1-x){\vec q_1}^2 + x{\vec q_2}^2}{(1-x){\vec q_1}^2} \right)
+ \frac{2({\vec q_1}{\vec \Delta})({\vec q_1}{\vec q_2})}{{\vec q_2}^2} \ln
\left( \frac{(1-x){\vec q_1}^2 + x{\vec q_2}^2}{(1-x){\vec q_1}^2} \right)
$$
\begin{equation}\label{AI2}
- \frac{({\vec q_1}{\vec q_2})^2}{{\vec q_2}^2} + \frac{(1-2x){\vec q_1}^2}
{(1-x)} \biggl\}.
\end{equation}

$$
I_3 = \int \mu^{-2\epsilon}\frac{d^{D-2}k_1}{(2\pi)^{D-1}}(D-2)\frac{((k_1 -
x\Delta)q_1)^2(q_1^2 - x\Delta^2 - 2k_1q_2)}{\kappa^2t} = \frac{2\Gamma(1 -
\epsilon)}{(4\pi)^{2 + \epsilon}}(2x^2(1-x)^2)
$$
$$
\times \biggl\{ {\vec q_1}^2 \left( \frac{1}{2\epsilon} - \frac{1}{2} + \frac
{({\vec q_1}{\vec q_2})^2}{{\vec q_1}^2{\vec q_2}^2} + \frac{1}{2} \ln \left(
\frac{x(1-x){\vec q_1}^2}{\mu^2} \right) \right) + \frac{{\vec q_1}^2((1-x)
{\vec q_1}^2 + x{\vec q_2}^2)}{x{\vec q_2}^2}\left( \frac{1}{2} - \frac{({\vec
q_1}{\vec q_2})^2}{{\vec q_1}^2{\vec q_2}^2} \right)
$$
\begin{equation}\label{AI3}
\times \ln \left( \frac{(1-x){\vec q_1}^2 + x{\vec q_2}^2}{(1-x){\vec q_1}^2}
\right) \biggr\}.
\end{equation}

$$
I_4 = \int \mu^{-2\epsilon}\frac{d^{D-2}k_1}{(2\pi)^{D-1}} (-(D-2)) \frac{(1-x)
(k_1q_1)^2(q_2^2 + x\Delta^2 + 2k_1q_2)}{\kappa t\tilde{t}} = \frac{2\Gamma(1 -
\epsilon)}{(4\pi)^{2 + \epsilon}}(x^2(1-x)^2)
$$
$$
\times \biggl\{ \frac{2x({\vec q_1}{\vec \Delta})^2}{((1-x){\vec q_1}^2 +
x{\vec q_2}^2)} \biggl( \frac{1}{\epsilon} + 1 + \ln\left( \frac{x(1-x){\vec
q_1}^2}{\mu^2} \right) \biggr) + \biggl( \frac{4x({\vec q_1}{\vec \Delta})^2}
{((1-x){\vec q_1}^2 + x{\vec q_2}^2)} - {\vec q_1}^2
$$
$$
+ \frac{(2{\vec q_1}{\vec q_2})}{{\vec q_2}^2} \left( 2{\vec q_1}^2 - ({\vec
q_1}{\vec q_2}) \right) + \frac{(1-x){\vec q_1}^4}{x{\vec q_2}^2} \left( \frac
{2({\vec q_1}{\vec q_2})^2}{{\vec q_1}^2{\vec q_2}^2} - 1 \right) \biggr) \ln
\left( \frac{(1-x){\vec q_1}^2 + x{\vec q_2}^2}{(1-x){\vec q_1}^2} \right) +
\biggl( {\vec q_1}^2{\vec q_2}^2
$$
$$
+ 4x(1-2x){\vec q_1}^4 - 2({\vec q_1}{\vec q_2})^2 \left( 1 + \frac{2x(1-x)
{\vec q_1}^2}{{\vec q_2}^2} \right) + 4x{\vec q_1}^2({\vec q_1}{\vec q_2})
\biggr) \frac{1}{\sqrt{{\vec q_2}^2({\vec q_2}^2 + 4x(1-x){\vec q_1}^2)}}
$$
\begin{equation}\label{AI4}
\times \ln \left( \frac{\sqrt{{\vec q_2}^2 + 4x(1-x){\vec q_1}^2} + \sqrt{{\vec
q_2}^2}}{\sqrt{{\vec q_2}^2 + 4x(1-x){\vec q_1}^2} - \sqrt{{\vec q_2}^2}}
\right) \biggr\}.
\end{equation}

$$
I_5 = \int \mu^{-2\epsilon}\frac{d^{D-2}k_1}{(2\pi)^{D-1}} \frac{(-2)xq_1^2}
{\kappa zt}\biggl[ \left( \frac{q_2^2}{2} \right)^2 - (1-x)q_1^2q_2^2 + \left(
\frac{D-2}{4} \right)(1-x)^2(q_1^2)^2 + (1-x)
$$
$$
\times \left( 2q_2^2 - (D-2)(1-x)q_1^2 \right) (k_1q_1) + (D-2)(1-x)^2(k_1q_1)^
2 \biggr] = \frac{2\Gamma(1 - \epsilon)}{(4\pi)^{2 + \epsilon}} \frac{2x{\vec
q_1}^2}{{\vec \Delta}^2}
$$
$$
\times \biggl\{ \frac{1}{((1-x){\vec q_1}^2 + x{\vec q_2}^2)}\frac{1}{\epsilon}
\left( \frac{x(1-x){\vec \Delta}^2}{\mu^2} \right)^{\epsilon} \biggl[ \frac
{{\vec q_2}^4}{4} - (1-x){\vec q_2}^2({\vec q_1}^2 - 2x({\vec q_1}{\vec
\Delta})) + \frac{(1 + \epsilon)}{2}
$$
$$
\times (1-x)^2({\vec q_1}^2 - 2x({\vec q_1}{\vec \Delta}))^2 \biggr] + \biggl[
-\frac{{\vec q_2}^4}{4} + (1-x){\vec q_2}^2({\vec q_1}^2 - 2x({\vec q_1}{\vec
\Delta})) - \frac{(1-x)^2}{2}({\vec q_1}^2 - 2x({\vec q_1}{\vec \Delta}))^2
$$
$$
- 2x(1-x)^3\frac{{\vec \Delta}^2}{{\vec q_2}^2}\left( ({\vec q_1}{\vec q_2})^2
- {\vec q_1}^2{\vec q_2}^2 \right) - \frac{(1-x)}{{\vec q_2}^2}\left( x{\vec
q_2}^2 + (1-x)({\vec q_1}^2 - {\vec \Delta}^2) \right) \biggl\{ ({\vec q_1}
{\vec q_2}) \left( {\vec q_2}^2 - (1-x) \right.
$$
$$
\left. \times ({\vec q_1}^2 - 2x({\vec q_1}{\vec \Delta})) \right)
+ \frac{(1-x)}{2{\vec q_2}^2} \left( x{\vec q_2}^2 + (1-x)({\vec q_1}^2 - {\vec
\Delta}^2) \right) \left( 2({\vec q_1}{\vec q_2})^2 - {\vec q_1}^2{\vec q_2}^2
\right) \biggl\} \biggr]
$$
$$
\times \frac{1}{\sqrt{\left( x{\vec q_2}^2 + (1-x)({\vec q_1}^2 + {\vec \Delta}
^2) \right)^2 - 4(1-x)^2{\vec q_1}^2{\vec \Delta}^2}}
$$
$$
\times \ln \left( \frac{x{\vec q_2}^2 + (1-x)({\vec q_1}^2 + {\vec \Delta}^2)
+ \sqrt{\left( x{\vec q_2}^2 + (1-x)({\vec q_1}^2 + {\vec \Delta}^2) \right)^2
- 4(1-x)^2{\vec q_1}^2{\vec \Delta}^2}}{x{\vec q_2}^2 + (1-x)({\vec q_1}^2 +
{\vec \Delta}^2) - \sqrt{\left( x{\vec q_2}^2 + (1-x)({\vec q_1}^2 + {\vec
\Delta}^2) \right)^2 - 4(1-x)^2{\vec q_1}^2{\vec \Delta}^2}} \right)
$$
$$
+ \biggl[ \frac{1}{((1-x){\vec q_1}^2 + x{\vec q_2}^2)}\biggl\{ \frac{{\vec
q_2}^4}{4} - (1-x){\vec q_2}^2({\vec q_1}^2 - 2x({\vec q_1}{\vec \Delta})) +
\frac{(1-x)^2}{2}({\vec q_1}^2 - 2x({\vec q_1}{\vec \Delta}))^2 \biggr\} +
\frac{(1-x)}{{\vec q_2}^2}
$$
$$
\times \biggl\{ ({\vec q_1}{\vec q_2}) \left( {\vec q_2}^2 -
(1-x)({\vec q_1}^2 - 2x({\vec q_1}{\vec \Delta})) \right) + \frac{(1-x)}
{2{\vec q_2}^2} \left( x{\vec q_2}^2 + (1-x){\vec q_1}^2 \right) \left(
2({\vec q_1}{\vec q_2})^2 - {\vec q_1}^2{\vec q_2}^2 \right) \biggl\} \biggr]
$$
\begin{equation}\label{AI5}
\times \ln \left( \frac{((1-x){\vec q_1}^2 + x{\vec q_2}^2)^2}{(1-x)^2{\vec
q_1}^2{\vec \Delta}^2} \right) - \frac{(1-x)^3{\vec \Delta}^2}{2{\vec q_2}^4}
\left( 2({\vec q_1}{\vec q_2})^2 - {\vec q_1}^2{\vec q_2}^2 \right) \ln \left(
\frac{{\vec q_1}^2}{{\vec \Delta}^2} \right) \biggr\}.
\end{equation}

$$
I_6 = \int \mu^{-2\epsilon}\frac{d^{D-2}k_1}{(2\pi)^{D-1}} \frac{1}{2t
\tilde{t}}\biggl[ -2q_1^2q_2^2 + \left( \frac{D-2}{4} \right)x(1-x)\left(
(\Delta^2 - q_2^2)^2 + (q_1^2)^2 \right) \biggr] = \frac{2\Gamma(1 - \epsilon)}
{(4\pi)^{2 + \epsilon}}
$$
$$
\times \frac{x(1-x)}{\sqrt{{\vec q_2}^2({\vec q_2}^2 + 4x(1-x){\vec q_1}^2)}}
\biggl[ -2{\vec q_1}^2{\vec q_2}^2 + \frac{x(1-x)}{2} \left( {\vec q_1}^4 +
({\vec \Delta}^2 - {\vec q_2}^2)^2 \right) \biggr]
$$
\begin{equation}\label{AI6}
\times \ln \left( \frac{\sqrt{{\vec q_2}^2 + 4x(1-x){\vec q_1}^2} + \sqrt{{\vec
q_2}^2}}{\sqrt{{\vec q_2}^2 + 4x(1-x){\vec q_1}^2} - \sqrt{{\vec q_2}^2}}
\right).
\end{equation}

$$
I_7 = \int \mu^{-2\epsilon}\frac{d^{D-2}k_1}{(2\pi)^{D-1}} \frac{x(1-x)^2q_1^2}
{2\kappa z}\biggl[ (D-2)(1-x)q_1^2 - (D-4)x(\Delta^2-q_2^2) \biggr] = \frac
{2\Gamma(1 - \epsilon)}{(4\pi)^{2 + \epsilon}}
$$
\begin{equation}\label{AI7}
\times \frac{x(1-x)^2{\vec q_1}^2}{{\vec \Delta}^2} \frac{1}{\epsilon} \left(
\frac{x(1-x){\vec \Delta}^2}{\mu^2} \right)^{\epsilon} \biggl[ -(1+\epsilon)(1
- x){\vec q_1}^2 + \epsilon x({\vec \Delta}^2 - {\vec q_2}^2) \biggr].
\end{equation}

$$
I_8 = \int \mu^{-2\epsilon}\frac{d^{D-2}k_1}{(2\pi)^{D-1}} \biggl( \frac
{x(1-x)q_1^2}{z} \biggr)^2\biggl[ \left( \frac{D-2}{4} \right) - (D-1)
x(1-x) \biggr] = \frac{2\Gamma(1 - \epsilon)}{(4\pi)^{2 + \epsilon}} \frac
{x(1-x){\vec q_1}^4}{{\vec \Delta}^2}
$$
\begin{equation}\label{AI8}
\times \left( \frac{x(1-x){\vec \Delta}^2}{\mu^2} \right)^{\epsilon} \left(
\frac{(1+\epsilon)}{2} - (3+2\epsilon)x(1-x) \right).
\end{equation}

$$
I_9 = \int \mu^{-2\epsilon}\frac{d^{D-2}k_1}{(2\pi)^{D-1}} \left( \frac{D-2}{2}
\right)(1-4x)\biggl( \frac{(1-x)q_1^2}{t} \biggr)^2 = \frac{2\Gamma(1 -
\epsilon)}{(4\pi)^{2 + \epsilon}} x(1-x)(1-4x){\vec q_1}^2
$$
\begin{equation}\label{AI9}
\times (1+\epsilon)\left( \frac{x(1-x){\vec q_1}^2}{\mu^2} \right)^{\epsilon}.
\end{equation}
The integrals $I_7$, $I_8$ and $I_9$ are calculated exactly.

$$
I_{10} = \int \mu^{-2\epsilon}\frac{d^{D-2}k_1}{(2\pi)^{D-1}} \frac{(1-x)
(2k_1q_1 - q_1^2)q_1^2q_2^2}{2tk_1^2k_2^2} = \frac{2\Gamma(1 - \epsilon)}
{(4\pi)^{2 + \epsilon}} \mu^{-2\epsilon} \frac{(1-x){\vec q_1}^2{\vec q_2}^2}
{2}
$$
$$
\times \biggl[ -\frac{1}{\epsilon} \frac{(x(1-x){\vec q_1}^2)^{\epsilon}}
{(x{\vec q_2}^2 + (1-x){\vec \Delta}^2)} + \frac{\Gamma^2(\epsilon)}
{\Gamma(2\epsilon)} \left( \frac{1}{(x{\vec q_2}^2 + (1-x){\vec \Delta}^2)^
{1-\epsilon}} - \frac{1}{({\vec \Delta}^2)^{1-\epsilon}} \right)
$$
\begin{equation}\label{AI10}
- \frac{2\epsilon ({\vec \Delta}^2)^{\epsilon}}{(x{\vec q_2}^2 + {\vec \Delta}^
2)} \int_{{\vec \Delta}^2/(x{\vec q_2}^2 + {\vec \Delta}^2)}^1 \frac{dz}{z}
\ln(1-z) \biggr].
\end{equation}

$$
I_{11} = \int \mu^{-2\epsilon}\frac{d^{D-2}k_1}{(2\pi)^{D-1}} \frac{(-x)q_2^2
(q_1^2)^2}{2\kappa tk_1^2} = \frac{2\Gamma(1 - \epsilon)}{(4\pi)^{2 +
\epsilon}} \frac{x(1-x){\vec q_1}^2{\vec q_2}^2}{2} \biggl[ \frac{\Gamma^2
(\epsilon)}{\Gamma(2\epsilon)} \frac{(x^2{\vec \Delta}^2/\mu^2)^{\epsilon}}
{x{\vec \Delta}^2}
$$
$$
+ \frac{(2({\vec q_1}{\vec \Delta}) - {\vec \Delta}^2)}{{\vec \Delta}^2(
x{\vec q_2}^2 + (1-x){\vec q_1}^2 )} \biggl( \frac{1}{\epsilon} + \ln \left(
\frac{x((1-x){\vec q_1}^2 + x{\vec q_2}^2)^2}{(1-x){\vec q_1}^2\mu^2} \right)
\biggr) - \frac{2({\vec \Delta}({\vec q_1} - x{\vec \Delta}))}{{\vec \Delta}^2
({\vec q_1} - x{\vec \Delta})^2}
$$
\begin{equation}\label{AI11}
\times \ln \left( \frac{(1-x){\vec q_1}^2 + x{\vec q_2}^2}{x(1-x){\vec
\Delta}^2} \right) \biggr].
\end{equation}

$$
I_{12} = \int \mu^{-2\epsilon}\frac{d^{D-2}k_1}{(2\pi)^{D-1}} \frac{(q_1^2)^2
(q_2^2)^2}{4t\tilde{t}k_1^2k_2^2} = \frac{1}{2}\biggl[ -I_{10} + \frac
{2\Gamma(1 - \epsilon)}{(4\pi)^{2 + \epsilon}}(1-x){\vec q_1}^2{\vec q_2}^2
$$
$$
\times \biggl\{ \frac{\left( {\vec q_2}^2({\vec \Delta}({\vec q_1} +
{\vec q_2})) - 2x({\vec q_1}{\vec q_2})({\vec q_2}{\vec \Delta}) \right)}
{{\cal D}(x){\vec \Delta}^2}\ln\left( \frac{(1-x){\vec q_2}^2 + x{\vec
\Delta}^2}{x{\vec q_1}^2} \right)
$$
$$
- \frac{{\vec q_2}^4\left( {\vec q_2}^2({\vec \Delta}({\vec q_1} + {\vec q_2}))
- 4x(1-x)(2{\vec q_1}^2{\vec q_2}^2 - ({\vec q_1}{\vec q_2})^2) \right)}{{\cal
D}(x){\cal D}(1-x)\sqrt{{\vec q_2}^2({\vec q_2}^2 + 4x(1-x){\vec q_1}^2)}}
\ln\left( \frac{\sqrt{{\vec q_2}^2 + 4x(1-x){\vec q_1}^2} + \sqrt{{\vec
q_2}^2}}{\sqrt{{\vec q_2}^2 + 4x(1-x){\vec q_1}^2} - \sqrt{{\vec q_2}^2}}
\right) \biggr\} \biggr]
$$
\begin{equation}\label{AI12}
+ \frac{1}{2}\biggl[ x \leftrightarrow (1 - x)
\biggr],\ \ \ \ \ \ \ {\cal D}(z) = \frac{\left( ({\vec q_2}^2 - 2z{\vec q_1}
{\vec q_2}){\vec q_1} + {\vec q_1}^2{\vec q_2} \right)^2}{{\vec q_1}^2}.
\end{equation}

\begin{equation}\label{AI13}
I_{13} = \int \mu^{-2\epsilon}\frac{d^{D-2}k_1}{(2\pi)^{D-1}} \frac{(-x)q_1^2
q_2^2}{2(1-x)\kappa k_1^2} = \frac{2\Gamma(1 - \epsilon)}{(4\pi)^{2 +
\epsilon}} \frac{{\vec q_1}^2{\vec q_2}^2}{2{\vec \Delta}^2} \frac{\Gamma^2
(\epsilon)}{\Gamma(2\epsilon)} \left( \frac{x^2{\vec \Delta}^2}{\mu^2}
\right)^{\epsilon}.
\end{equation}
$I_{13}$ is calculated exactly.

$$
I_{14} = \int \mu^{-2\epsilon}\frac{d^{D-2}k_1}{(2\pi)^{D-1}} \frac{x^2q_1^2
q_2^2}{2k_1^2z} = \frac{2\Gamma(1 - \epsilon)}{(4\pi)^{2 + \epsilon}} \frac
{{\vec q_1}^2{\vec q_2}^2}{2{\vec \Delta}^2} \left( \frac{{\vec \Delta}^2}
{\mu^2} \right)^{\epsilon} x^2 \int_0^1 \frac{dz}{\left( xz(1-xz) \right)^
{1-\epsilon}} \approx \frac{2\Gamma(1 - \epsilon)}{(4\pi)^{2 + \epsilon}}
$$
\begin{equation}\label{AI14}
\times \frac{{\vec q_1}^2{\vec q_2}^2}{2{\vec \Delta}^2} \left( \frac{{\vec
\Delta}^2}{\mu^2} \right)^{\epsilon} x \biggl[ \frac{\Gamma^2(\epsilon)}{\Gamma
(2\epsilon)} - \frac{(1-x)^{\epsilon}}{\epsilon} + \ln x + \frac{\epsilon}{2}
\ln x \ln (x(1-x)^2) - 2\epsilon\int_0^{1-x} \frac{dz}{z} \ln (1-z) \biggr].
\end{equation}
The first relation here is exact and in the second approximate relation we
have made the expansion in $\epsilon$ with accuracy needed.

\begin{equation}\label{AI15}
I_{15} = \int \mu^{-2\epsilon}\frac{d^{D-2}k_1}{(2\pi)^{D-1}} \frac{q_1^2q_2^2}
{2k_1^2k_2^2} = \frac{2\Gamma(1 - \epsilon)}{(4\pi)^{2 + \epsilon}} \frac
{{\vec q_1}^2{\vec q_2}^2}{2{\vec \Delta}^2} \frac{\Gamma^2(\epsilon)}{\Gamma
(2\epsilon)} \left( \frac{{\vec \Delta}^2}{\mu^2} \right)^{\epsilon}.
\end{equation}
The last relation is exact.

\newpage

{\bf Appendix II.}

Here we calculate the integrals appearing in the total cross section
(\ref{19}). The first six integrals are calculated exactly without any
problems and the result is

$$
J_1 + ... + J_6 = \int_{\delta_R}^{1-\delta_R}\frac{dx}{x(1-x)} \int
\mu^{-2\epsilon}\frac{d^{D-2}k_1}{(2\pi)^{D-1}} \biggl\{ (1 + \epsilon)
(1-x)^2 \frac{4(k_1q_1)^2 + (1-4x)(q_1^2)^2}{t^2}
$$
$$
+ \frac{x(1-x)q_1^2}{\kappa z} \biggl[ 2q_2^2 + (1 + \epsilon) \left( 2(1-x)
(k_1q_1) - x(1-x)q_1^2 - k_2^2 \right) - \epsilon x(1-x) (\Delta^2 - q_2^2)
\biggr]
$$
$$
+ \left( \frac{x(1-x)q_1^2}{z} \right)^2 \biggl[ \frac{(1 + \epsilon)}{2} -
(3 + 2\epsilon)x(1-x) \biggr] + \biggl( \frac{-xq_1^2q_2^2}{2(1-x)\kappa k_1^2}
+ \frac{x^2q_1^2q_2^2}{2k_1^2z} + \frac{xq_1^2q_2^2}{k_1^2k_2^2} \biggr)
\biggr\}
$$
$$
= \frac{\Gamma(1-\epsilon)}{(4\pi)^{2+\epsilon}}\frac{2\Gamma^2(1 + \epsilon)}
{\epsilon\Gamma(1+2\epsilon)} \biggl[ \biggl( 4\ln(1/\delta_R) + \psi(1) +
\psi(1+\epsilon) - 2\psi(1+2\epsilon) - \frac{(11+7\epsilon)}{2(1+2\epsilon)
(3+2\epsilon)} \biggr)
$$
\begin{equation}\label{AII1}
\times \left( \frac{{\vec \Delta}^2}{\mu^2} \right)^{\epsilon} \frac{{\vec
q_1}^2{\vec q_2}^2}{{\vec \Delta}^2} - \frac{(1 + \epsilon)}{(1 + 2\epsilon)
(3 + 2\epsilon)} \left( \frac{{\vec q_1}^2}{\mu^2} \right)^{\epsilon} {\vec
q_1}^2 \biggr].
\end{equation}

All others integrals don't contain of the dependence on $\delta_R$ at $\delta_R
\rightarrow 0$, therefore we will put $\delta_R = 0$ below. The next two
integrals aren't singular at $| {\vec \Delta} | \rightarrow 0$ and we can
calculate them with accuracy up to $const(\epsilon)$. Besides that, using
the following change of integration variables, corresponding to left-right
symmetry \cite{6},
\begin{equation}\label{AII2}
{\vec k_1} \leftrightarrow {\vec k_2},\ \ \ \ \ x \leftrightarrow \frac{x{\vec
k_2}^2}{((1-x){\vec k_1}^2 + x{\vec k_2}^2)},
\end{equation}
we obtain
$$
J_8 = \int_0^1 \frac{dx}{x(1-x)} \int \mu^{-2\epsilon}\frac{d^{D-2}k_1}
{(2\pi)^{D-1}} \frac{(-x)q_1^2k_2^2\left( (1 + \epsilon)k_2^2 -2q_2^2 \right)}
{\kappa zt} =
$$
\begin{equation}\label{AII3}
\biggl[ \int_0^1 \frac{dx}{x(1-x)} \int \mu^{-2\epsilon}\frac
{d^{D-2}k_1}{(2\pi)^{D-1}} \frac{(-x)q_2^2 \left( (1 + \epsilon)k_1^2 -2q_1^2
\right)}{\kappa t} \biggr] (q_1 \leftrightarrow -q_2),
\end{equation}
and therefore $J_8$ has the same structure as $J_7$ and it is convenient to
consider these two integrals together. After rather long calculation we get
$$
J_7 + J_8 = \int_0^1 \frac{dx}{x(1-x)} \int \mu^{-2\epsilon}\frac{d^{D-2}k_1}
{(2\pi)^{D-1}} \biggl\{ \frac{1}{\kappa t}\biggl[ -2(1-x)q_1^2q_2^2 + (1 +
\epsilon) (1-x) (q_1^2)^2 + \frac{(1 + \epsilon)}{2}
$$
$$
\times x(1-x)\left( 2(1-x)q_1^2(\Delta^2 - q_2^2) - x(q_1^2)^2 - x(\Delta^2 -
q_2^2)^2 \right) -2(1 + \epsilon)(2-x)(1-x)q_1^2(k_1q_1)
$$
$$
+ 2(1 + \epsilon)(1-x)\left( (k_1q_1)^2 + ((k_1 - x\Delta)q_1)^2 \right)
+ (1 + \epsilon)q_1^2k_2^2 \biggr] - \frac{xq_1^2k_2^2\left( (1 + \epsilon)
k_2^2 -2q_2^2 \right)}{\kappa zt} \biggr\} =
$$
$$
\frac{\Gamma(1-\epsilon)}{(4\pi)^{2+\epsilon}}\frac{2\Gamma^2(1 + \epsilon)}
{\epsilon\Gamma(1+2\epsilon)} \frac{(1 + \epsilon)}{(1 + 2\epsilon)(3 +
2\epsilon)} \biggl( \left( \frac{{\vec q_1}^2}{\mu^2} \right)^{\epsilon} -
\left( 1 + \frac{\epsilon}{2} \right) \left( \frac{{\vec q_2}^2}{\mu^2} \right)
^{\epsilon} \biggl) {\vec q_1}^2 - \frac{\Gamma(1-\epsilon)}{(4\pi)^{2 +
\epsilon}} \biggl[ \frac{2}{3}\biggl( {\vec q_1}^2 - 2
$$
$$
\times \frac{({\vec q_1}{\vec q_2})^2}{{\vec q_2}^2} \biggr) - \frac{{\vec
q_1}^2}{3}\left( \frac{1}{\epsilon} + \ln \left( \frac{{\vec q_1}^2}{\mu^2}
\right) - \frac{2}{3} \right) + (2{\vec q_1}{\vec \Delta}) \left( \frac{2}{3}
\ln \left( \frac{{\vec q_1}^2}{{\vec q_2}^2} \right) + 1 \right) - 2{\vec
\Delta}^2 + \frac{2}{3} \frac{{\vec q_1}^2(6{\vec q_2}^2 - {\vec \Delta}^2)}
{({\vec q_1}^2 - {\vec q_2}^2)}
$$
$$
\times \ln \left( \frac{{\vec q_1}^2}{{\vec q_2}^2} \right) + \frac
{{\vec \Delta}^2{\vec q_2}^2}{({\vec q_1}^2 - {\vec q_2}^2)^3} \left(
{\vec q_1}^2({\vec q_1}^2 - 3{\vec q_2}^2) \ln \left( \frac{{\vec q_1}^2}
{{\vec q_2}^2} \right) + 2{\vec q_2}^2({\vec q_1}^2 - {\vec q_2}^2) \right)
+ \frac{2}{3({\vec q_1}^2 - {\vec q_2}^2)^3} \biggl( {\vec q_1}^2(3{\vec q_2}^2
- {\vec q_1}^2)
$$
\begin{equation}\label{AII4}
- ({\vec q_1}{\vec q_2})(3{\vec q_1}^2 + {\vec q_2}^2) + 2{\vec q_1}^2\frac
{({\vec q_1}{\vec q_2})^2}{{\vec q_2}^2} \biggr) \left( 2{\vec q_2}^4 \ln
\left( \frac{{\vec q_1}^2}{{\vec q_2}^2} \right) - 2{\vec q_2}^2({\vec q_1}^2
- {\vec q_2}^2) + ({\vec q_1}^2 - {\vec q_2}^2)^2 \right) \biggr].
\end{equation}

Using the change of integration variables (\ref{AII2}) one can get
\begin{equation}\label{AII5}
J_{10} = J_9 (q_1 \leftrightarrow -q_2),
\end{equation}
and it is enough to calculate only one of these integrals. For the sum
of $J_9$ and $J_{10}$ we obtain
$$
J_9 + J_{10} = \int_0^1 \frac{dx}{x(1-x)} \int \mu^{-2\epsilon} \frac{d^{D-2}
k_1}{(2\pi)^{D-1}} \biggl\{ - \frac{xq_1^2(q_2^2)^2}{2\kappa zt} - \frac
{xq_2^2(q_1^2)^2}{2\kappa tk_1^2} \biggr\}
$$
\begin{equation}\label{AII6}
= \frac{\Gamma(1-\epsilon)}
{(4\pi)^{2+\epsilon}} \frac{\Gamma^2(1 + \epsilon)}{\Gamma(1+2\epsilon)}
\frac{2}{\epsilon^2} \left( \frac{{\vec \Delta}^2}{\mu^2} \right)^{\epsilon}
\frac{{\vec q_1}^2{\vec q_2}^2}{{\vec \Delta}^2}.
\end{equation}

$$
J_{11} = \int_0^1 \frac{dx}{x(1-x)} \int \mu^{-2\epsilon} \frac{d^{D-2}k_1}
{(2\pi)^{D-1}} \frac{(1-x)(2(k_1q_1) - q_1^2)q_1^2q_2^2}{2tk_1^2k_2^2} =
-\frac{\Gamma(1-\epsilon)}{(4\pi)^{2+\epsilon}} \frac{\Gamma^2(1 + \epsilon)}
{\Gamma(1+2\epsilon)}
$$
\begin{equation}\label{AII7}
\times \biggl[ \frac{1}{\epsilon^2} + \frac{1}{\epsilon} \ln
\left( \frac{{\vec q_1}^2{\vec q_2}^2}{{\vec \Delta}^4} \right) + \frac{1}{2}
\ln^2 \left( \frac{{\vec q_1}^2}{{\vec q_2}^2} \right) + 4\epsilon\psi^
{\prime\prime}(1) \biggr] \left( \frac{{\vec \Delta}^2}{\mu^2} \right)^
{\epsilon}\frac{{\vec q_1}^2{\vec q_2}^2}{{\vec \Delta}^2}.
\end{equation}

$$
J_{12} = \int_0^1 \frac{dx}{x(1-x)} \int \mu^{-2\epsilon} \frac{d^{D-2}k_1}
{(2\pi)^{D-1}} \frac{1}{2t\tilde{t}}\biggl[ -2q_1^2q_2^2 + \frac{(1 +
\epsilon)}{2}x(1-x)\left( (\Delta^2 - q_2^2)^2 + (q_1^2)^2 \right)
$$
$$
- 4(1 + \epsilon)x(1-x)(k_1q_1)^2 \biggr] = \frac{\Gamma(1 - \epsilon)}
{(4\pi)^{2 + \epsilon}} \biggl[ \frac{{\vec q_1}^2}{3}\left( \frac{1}{\epsilon}
+ \ln \left( \frac{{\vec q_1}^2}{\mu^2} \right) - \frac{1}{6} \right) + \frac
{{\vec q_2}^2}{2} \biggr] - \frac{\Gamma(1 - \epsilon)}{(4\pi)^{2 + \epsilon}}
$$
$$
\times \biggl\{ \frac{3}{8}\frac{\left( 2{\vec q_1}^2{\vec q_2}^2 - ({\vec q_1}
{\vec q_2})^2 \right)}{{\vec q_1}^2{\vec q_2}^2}({\vec q_1}^2 + {\vec q_2}^2) +
\frac{\left( 2{\vec q_1}^2{\vec q_2}^2 - 3({\vec q_1}{\vec q_2})^2 \right)}
{16{\vec q_1}^2{\vec q_2}^2}({\vec q_1}^2 - {\vec q_2}^2) \ln \left( \frac
{{\vec q_1}^2}{{\vec q_2}^2} \right) + \biggl[ 4{\vec q_1}^2{\vec q_2}^2 +
\frac{({\vec q_1}^2 - {\vec q_2}^2)^2}{4}
$$
\begin{equation}\label{AII8}
+ \left( 2{\vec q_1}^2{\vec q_2}^2 -3{\vec q_1}^4 - 3{\vec q_2}^4 \right)\frac
{\left( 2{\vec q_1}^2{\vec q_2}^2 - ({\vec q_1}{\vec q_2})^2 \right)}{16{\vec
q_1}^2{\vec q_2}^2} \biggr] \int_0^{\infty} \frac{dx}{({\vec q_1}^2 + x^2{\vec
q_2}^2)} \ln \left| \frac{1+x}{1-x} \right| \biggl\}.
\end{equation}

$$
J_{13} = \int_0^1 \frac{dx}{x(1-x)} \int \mu^{-2\epsilon} \frac{d^{D-2}k_1}
{(2\pi)^{D-1}} \frac{(q_1^2)^2(q_2^2)^2}{4t\tilde{t} k_1^2k_2^2} =
\frac{\Gamma(1-\epsilon)}{(4\pi)^{2+\epsilon}} \Biggl\{ \frac{{\vec q_1}^2
{\vec q_2}^2}{{\vec \Delta}^2}\left( \frac{{\vec \Delta}^2}{\mu^2} \right)^
{\epsilon}\frac{\Gamma^2(1+\epsilon)}{\Gamma(1 + 2\epsilon)} \Biggl[ \frac{1}
{\epsilon^2}
$$
$$
+ \frac{1}{\epsilon}\ln\left( \frac{{\vec q_1}^2{\vec q_2}^2}{{\vec \Delta}^4}
\right) + \frac{1}{2}\ln^2\left( \frac{{\vec q_1}^2}{{\vec q_2}^2} \right) +
4\epsilon\psi^{\prime\prime}(1) \Biggr] + \frac{2{\vec q_1}^2{\vec q_2}^2
({\vec \Delta}(\vec q_1 + \vec q_2))}{{\vec \Delta}^2(\vec q_1 + \vec q_2)^2}
\Biggl[ \ln\left( \frac{{\vec q_1}^2}{{\vec q_2}^2} \right)\ln\left( \frac
{{\vec q_1}^2{\vec q_2}^2}{({\vec q_1}^2 + {\vec q_2}^2)^2} \right)
$$
$$
+ L\left( 1-\frac{{\vec \Delta}^2}{{\vec q_2}^2} \right) - L\left( 1-\frac
{{\vec \Delta}^2}{{\vec q_1}^2} \right) + L\left( -\frac{{\vec q_1}^2}{{\vec
q_2}^2} \right) - L\left( -\frac{{\vec q_2}^2}{{\vec q_1}^2} \right) \Biggr]
+ 2{\vec q_1}^2{\vec q_2}^2 \Biggl[ \Biggl( \int_0^1 \frac{dt}{({\vec q_2}^2t^2
- 2({\vec q_1}{\vec q_2})t + {\vec q_1}^2)}
$$
\begin{equation}\label{AII9}
\times \Biggl( \frac{({\vec q_2}{\vec \Delta})}{{\vec \Delta}^2} - \frac{{\vec
q_2}^2({\vec \Delta}(\vec q_1 + \vec q_2))}{{\vec \Delta}^2(\vec q_1 + \vec
q_2)^2}(1+t) \Biggr) \ln \left( \frac{{\vec q_2}^2t(1-t)}{{\vec q_1}^2(1-t)
+ {\vec \Delta}^2t} \right) \Biggr) + \Biggl( \vec q_1 \leftrightarrow -\vec
q_2 \Biggr) \Biggr] \Biggr\},
\end{equation}
where
\begin{equation}\label{AII10}
L(x) = \int_0^x \frac{dt}{t} \ln(1-t).
\end{equation}


\newpage

\begin{thebibliography}{99}
\bibitem{1}
V.S. Fadin, E.A. Kuraev and L.N.Lipatov, Phys. Lett. {\bf B 60} (1975) 50;
E.A.Kuraev, L.N. Lipatov and V.S. Fadin, Zh. Eksp. Teor. Fiz. {\bf 71} (1976)
840 [Sov. Phys. JETP {\bf 44} (1976) 443]; {\bf 72} (1977) 377 [{\bf 45} (1977)
199]. Ya.Ya. Balitskii and L.N. Lipatov, Sov. J. Nucl. Phys. {\bf 28} (1978)
822.

\bibitem{2}  A.J. Askew, J. Kwiecinski, A.D. Martin and P.J. Sutton,
Phys. Rev. {\bf D 49} (1994) 4402; A.J. Askew, K. Golec-Biernat, J.
Kwiecinski, A.D. Martin and P.J. Sutton, Phys. Lett. {\bf B 325} (1994) 212.

\bibitem{3}  L. N. Lipatov and V. S. Fadin, Sov. Phys. JETP Lett.
${\bf {49}}$ (1989) 352;\\Sov. J. Nucl. Phys. ${\bf {50}}$
(1989) 712.

\bibitem{4}
V.S. Fadin and L.N. Lipatov, Nucl. Phys. {\bf B 406} (1993) 259. V.S. Fadin,
R. Fiore and A. Quartarolo, Phys. Rev. {\bf D 50} (1994) 5893. V.S. Fadin,
R. Fiore, M.I. Kotsky, preprint BUDKERINP/96-52, CS-TH 4/96 (1996), to be
published in Phys. Lett. {\bf B}.

\bibitem{5}
V.S. Fadin, R. Fiore and A. Quartarolo, Phys. Rev. {\bf D 53} (1996) 2729;
V.S. Fadin, Zh. Eksp. Teor. Fiz. Pis'ma {\bf 61} (1995) 342;
M.I.Kotsky and V.S. Fadin, Yadernaya Fizika {\bf 59}(6) (1996) 1;
V.S. Fadin, R. Fiore and M.I. Kotsky, Phys. Lett. {\bf B 359} (1995) 181;
V.S. Fadin, R. Fiore and M.I. Kotsky, Phys. Lett. {\bf B 387 (1996) 593}.

\bibitem{6}
V.S.Fadin,  L.N.Lipatov, Nucl. Phys. {\bf B 477} (1996) 767.

\bibitem{7}
V.S.Fadin, Talk given at the Internaional Conference "Deep Inelastic
Scattering 96", April 14-19, 1996, Rome, Italy.

\bibitem{8}
G. Camici and M. Ciafaloni, preprint DFF 250/6/96, HEP-PH 9606427 (1996).


\end{thebibliography}
\end{document}